\documentstyle[12pt,oldlfont]{article}

\font\tenrm=cmr10

\begin{document}
\renewenvironment{thebibliography}[1]
  { \begin{list}{\arabic{enumi}.}
    {\usecounter{enumi} \setlength{\parsep}{0pt}
     \setlength{\itemsep}{3pt} \settowidth{\labelwidth}{#1.}
     \sloppy
    }}{\end{list}}

\parindent=1.5pc

\renewcommand{\thefootnote}{\fnsymbol{footnote} }

\newcommand{\s}{\\ \vspace*{-2mm} }
\newcommand{\nn}{\noindent}
\newcommand{\non}{\nonumber}
\newcommand{\ee}{e^+ e^-}
\newcommand{\ra}{\rightarrow}
\newcommand{\lra}{\longrightarrow}
\newcommand{\beq}{\begin{eqnarray}}
\newcommand{\eeq}{\end{eqnarray}}
\newcommand{\tb}{{\rm tg} \beta}

% ************************* preprint ******************************

\hspace*{10cm} UdeM--GPP--TH--04 \\

\vspace*{-3mm}

\begin{center}{{\bf THE SEARCH FOR HIGGS BOSONS: A SUMMARY\footnote{Talk given
at the
MRST--1994 Meeting ``What next? Exploring the Future of High--Energy Physics",
11--13 May 1994, Montr\'eal, Canada.}}\s
\vglue 0.5cm
{A. DJOUADI}\\
\baselineskip=14pt
{\it Groupe de Physique des Particules, Universit\'e de Montr\'eal,}\\
\baselineskip=14pt
{\it Case 6128A, H3C 3J7, Montr\'eal P.Q., Canada.}\s
%=========================================================================
\vglue 0.5cm
{\tenrm ABSTRACT}}
\end{center}
%  \vglue 0.2cm
{\rightskip=3pc
 \leftskip=3pc
 \tenrm\baselineskip=12pt
 \noindent
I summarize the prospects for discovering Higgs bosons at future $pp$ and
$\ee$ colliders both in the Standard Model and its Minimal Supersymmetric
extension.
\vglue 0.8cm}
%
%=========================================================================
%   BODY OF PAPER BEGINS HERE
%
{\bf\noindent 1. Introduction}
\vglue 0.2cm
\baselineskip=14pt

In order to accomodate the well--established electromagnetic and weak
interaction phenomena, the existence of at least one isodoublet scalar field to
generate fermion and weak gauge bosons masses is required. The Standard Model
makes use of one isodoublet: three Goldstone bosons among the four degrees of
freedom are absorbed to build up the longitudinal components of the $W^\pm,Z$
gauge bosons; one degree of freedom is left over, corresponding to a physical
scalar particle, the Higgs boson.$^{1,2}$ The discovery of this particle is
the ultimate test of the Standard Model.

In the Standard Model (SM), the mass of the Higgs particle is a free parameter.
The only available information is the upper limit $M_H> 63.8$ GeV established
at LEP1;$^3$ this limit can be raised to 80--90 GeV at LEP2. However,
interesting constraints can be derived from assumptions on the energy range
within which the model is valid before perturbation theory breaks down and new
phenomena would emerge. First, if the Higgs mass were larger than $\sim$ 1 TeV,
the $W$ and $Z$ gauge bosons would interact strongly with each other to ensure
unitarity in their scattering at high energies. Second, the quartic Higgs
self--coupling, which at the scale $M_H$ is fixed by $M_H$ itself, grows
logarithmically with the energy scale. If $M_H$ is small, the energy cut--off
$\Lambda$ at which the coupling grows beyond any bound and new phenomena should
occur, is large; conversely, if $M_H$ is large, $\Lambda$ is small. The
condition $M_H < \Lambda$ sets an upper limit on the Higgs mass in the SM;
lattice analyses lead to an estimate of about 630 GeV for this limit. Requiring
that the SM be extended to the GUT scale, $\Lambda> 10^{15}$ GeV, and including
the effect of top quark loops on the running coupling, the Higgs mass should be
less than 180 to 200 GeV.

However, there are two problems that one has to face when trying to extend the
SM to $\Lambda_{\rm GUT}$. The first one is the so--called hierarchy or
naturalness problem: the Higgs boson tends to acquire a mass of the order of
these large scales [the radiative corrections to $M_H$ are quadratically
divergent]; the second one is that the simplest GUTs predict a value for
$\sin^2\theta_W$ that is incompatible with the measured one $\simeq~0.23$.

Low energy supersymmetry solves these two problems at once: supersymmetric
particle loops cancel the quadratic divergences and contribute to the running
of the gauge coupling constants, correcting the small discrepancy to the
observed value of $\sin^2\theta_W$; see Ref.~[4] for a review.

The minimal supersymmetric extension of the Standard Model (MSSM)$^{1,4}$
requires the existence of two isodoublets of Higgs fields, to cancel anomalies
and to give mass separately to up and down--type fermions. Three
neutral, $h/H$(CP=+), $A$(CP=--) and a pair of charged scalar particles,
$H^\pm$, are introduced by this extension of the Higgs sector. Besides the four
masses, two additional parameters define the properties of these particles: a
mixing angle $\alpha$ in the neutral CP--even sector and the ratio of the two
vacuum expectation values $\tb$, which from GUT restrictions is assumed in the
range $1 < \tb <m_t/m_b$. Supersymmetry leads to several relations among these
parameters and only two of them are in fact independent. These relations impose
a strong hierarchical structure of the mass spectrum, $M_h<M_Z, M_A<M_H$ and
$M_W<M_{H^\pm}$, which however is broken by radiative corrections if the top
quark mass is large.$^4$ For instance, the upper bound on the mass of
the lightest Higgs boson $h$ is shifted from the tree level value $M_Z$ to
$\sim 140$ GeV for $m_t<200$ GeV. The masses of the heavy neutral and charged
Higgs particles can be expected, with a high probability, in the range of the
electroweak symmetry breaking scale. Some of these features are not specific to
the minimal extension and are expected to be realized also in more general SUSY
models. For instance, a light Higgs boson with a mass below ${\cal O}$(200 GeV)
is quite generally predicted by SUSY theories.$^1$

The search for these Higgs particles will be a major goal of the next
generation of colliders. In the following, I will briefly discuss the discovery
potential of the pp collider LHC$^{5,6}$ with a c.m. energy of $\sim 14$ TeV as
well as a future ${\rm \ee}$ linear collider$^{7,8}$ with a c.m. energy in the
range of 300 to 500 GeV. A more detailed discussion and a complete set of
references can be found in Refs.~[4--8] or in the review of Ref.~[2]. \\

{\bf\noindent 2. Couplings and Decay Modes} \\
\vglue 0.2cm
\baselineskip=14pt

In the SM, the profile of the Higgs particle is uniquely determined once
$M_H$ is fixed. The decay width, the branching ratios and the production cross
sections are given by the strength of the Yukawa couplings to fermions and
gauge bosons, the scale of which is set by the masses of these particles. To
discuss the Higgs decay modes, it is convenient to divide the Higgs mass into
two ranges: the ``low mass" range $M_H<140$ GeV and the ``high mass" range
$M_H>140$ GeV.

In the ``low mass" range, the Higgs boson decays into a large variety of
channels. The main decay mode is by far the decay into $b\bar{b}$ pairs with a
branching ratio of $\sim 90\%$ followed by the decays into $c\bar{c}$ and
$\tau^+\tau^-$ pairs with a branching ratio of $\sim 5\%$. Also of
significance, the top--loop mediated Higgs decay into gluons, which for $M_H$
around 120 GeV occurs at the level of $\sim 5\%$. The top and $W$--loop
mediated $\gamma\gamma$ and $Z \gamma$ decay modes are very rare the branching
ratios being of ${\cal O }(10^{-3})$; however these decays lead to clear
signals and are interesting being sensitive to new heavy particles.

In the ``high mass" range, the Higgs bosons decay into $WW$ and $ZZ$ pairs,
with one of the gauge bosons being virtual below the threshold. Above the $ZZ$
threshold, the Higgs boson decays almost exclusively into these channels with a
branchings ratio of 2/3 for $WW$ and 1/3 for $ZZ$. The opening of the
$t\bar{t}$ channel does not alter significantly this pattern, since for large
Higgs masses, the $t\bar{t}$ decay width rises only linearly with $M_H$ while
the decay widths to $W$ and $Z$ bosons grow with $M_H^3$.

In the low mass range, the Higgs boson is very narrow $\Gamma_H<10$ MeV, but
the width becomes rapidly wider for masses larger than 140 GeV, reaching 1 GeV
at the $ZZ$ threshold; the Higgs decay width cannot be measured directly in the
mass range below 250 GeV.

In the MSSM, since the lightest CP--even Higgs boson $h$ is likely to be
the particle which will be discovered first, an attractive choice of the two
input parameters is the set $(M_h, \tb $) [with $\tb$ parametrizing the
production cross sections]. Once these two parameters [as well as the top quark
mass and the associated squark masses which enter through radiative
corrections] are specified, all other masses and the angle $\alpha$ can be
derived.

The couplings of the various neutral Higgs bosons [collectively denoted by
$\Phi$] to fermions and gauge bosons will in general strongly depend on the
angles $\alpha$ and $\beta$; normalized to the SM Higgs couplings, they are
given by
\begin{center}
\begin{tabular}{|c|c|c|c|c|} \hline
$\ \ \ \Phi \ \ \ $ &$ g_{\Phi \bar{u}u} $	& $ g_{\Phi \bar{d} d} $ &
$g_{ \Phi VV} $ \\ \hline
$h$  & \ $\; \cos\alpha/\sin\beta	\; $ \ & \ $ \;	-\sin\alpha/
\cos\beta \; $ \ & \ $ \; \sin(\beta-\alpha) \;	$ \ \\
 $H$  & \	$\; \sin\alpha/\sin\beta \; $ \	& \ $ \; \cos\alpha/
\cos\beta \; $ \ & \ $ \; \cos(\beta-\alpha) \;	$ \ \\
$A$  & \ $\; 1/ \tb \; $\ & \ $	\; \tb \; $ \	& \ $ \; 0 \; $	\ \\ \hline
\end{tabular}
\end{center}
\vspace{0.3cm}

The pseudoscalar has no tree level couplings to gauge bosons, and its
couplings to down (up) type fermions are (inversely) proportional to $\tb$.
For the CP--even Higgs bosons, the couplings to down (up) type fermions are
enhanced (suppressed) compared to the	SM Higgs couplings [$\tb>1$]. If $M_h$
is very close to its upper limit for a given value of $\tb$, the couplings
to fermions and gauge bosons are SM like. If all other Higgs bosons are
very heavy, it is very difficult to distinguish the Higgs sector
of the MSSM from the SM.

The lightest Higgs boson will decay mainly into fermion pairs since its
mass is smaller than $\sim$ 140 GeV. This is also the dominant decay mode of
the pseudoscalar $A$ which has no tree--level couplings to gauge bosons. For
values of $\tb$ larger than unity and for masses less than $\sim$ 140 GeV, the
main decay modes of the neutral Higgs bosons will be decays into $b \bar{b}$
and $\tau^+ \tau^-$ pairs; the branching ratios always being larger than $
\sim 90\%$ and $5\%$, respectively. The decays into $c\bar{c}$ and $gg$ are in
general strongly suppressed especially for large values of $\tb$. For high
masses, the top decay channels $H,A \rightarrow t\bar{t}$ open up; yet this
mode remains suppressed for large $\tb$.

If the mass is high enough, the heavy $H$ can in principle decay into $WW$
and $ZZ$ bosons but since the partial widths are proportional to $\cos^2(\beta-
\alpha)$, they are strongly suppressed. For the same reason, the cascade decay
of the pseudoscalar $ A \rightarrow	Zh$ is suppressed in general. The heavy
$H$ boson can also decay into two lighter Higgs bosons; but these modes are
restricted to very small domains in the parameter space. The branching ratios
of the $\gamma \gamma$ and $Z \gamma$ decays are smaller than in the SM; this
is due to the fact that the decays into $b\bar{b}$ are enhanced for $\tb>1$ and
the dominant $W$--loop contribution is suppressed (absent) in the case of the
CP--even (odd) Higgs bosons. Other possible channels are the decays into SUSY
particles: while sfermions are likely too heavy to affect Higgs decays, the
Higgs decays into charginos and neutralinos could eventually be important
since some of these particles are expected to have masses of ${\cal O}(M_Z$).

The	couplings of the charged Higgs particle	to fermions is a mixture of
scalar and pseudoscalar currents. The charged Higgs decays into fermions but
also, if allowed kinematically, into $hW$. Below the $tb$ and $Wh$ thresholds,
the charged Higgs particles will decay mostly into $\tau \nu_\tau$ and
$c\bar{s}$ pairs, the former being dominant for $\tb >1$.  For large
$M_{H^\pm}$ and $\tb$ values,	the top--bottom decay $H^+ \rightarrow
t\bar{b}$ becomes dominant.

Adding up the various decay modes, the width of all five Higgs bosons
remains very narrow, being of the order of 1 GeV even for large masses.\\

{\bf\noindent 3. Production at pp Colliders}
\vglue 0.2cm
\baselineskip=14pt

The main production mechanisms of neutral Higgs bosons at hadron colliders
are the following,$^{5,6}$
\begin{eqnarray}
\begin{array}{lccl}
(a) & \ \ {\rm gluon-gluon~fusion} & \ \ gg  \ \ \ra & H \nonumber \\
(b) & \ \ WW/ZZ~{\rm fusion}       & \ \ VV \  \ra &  H \nonumber \\
(c) & \ \ {\rm association~with}~W/Z & \ \ q\bar{q} \ \ \ra & V + H \nonumber
\\
(d) & \ \ {\rm association~with~}\bar{t}t & gg,q\bar{q}\ra & t\bar{t}+H
\nonumber
\end{array}
\end{eqnarray}

In the most interesting mass range, $80<M_H<250$ GeV, the dominant
production process is the gluon--gluon fusion mechanism, for which the
cross section is of order a few tens of pb. It is followed by the $WW/ZZ$
fusion processes with a cross section of a few pb; those of the associated
production with $W/Z$ or $t\bar{t}$ are an order of magnitude smaller. Note
that
for a luminosity of ${\cal L}=10^{33} (10^{34})$~cm$^{-2}$s$^{-1}$, $\sigma=$~1
pb would correspond to $10^{4}(10^{5})$ events per year.

Besides the errors due to the poor knowledge of the gluon distribution at
small $x$, the lowest order cross sections are affected by large uncertainties
due to higher order corrections. Including the next to leading QCD corrections,
the total cross sections can be defined properly: the scale at which one
defines
the strong coupling constant is fixed and the [generally non--negligible]
corrections are taken into account. The ``K--factors" for $WH/ZH$ production
[which can be inferred from the Drell--Yan production of weak bosons] and the
$VV$ fusion mechanisms are small, increasing the total cross sections by
$\sim$ 20 and 10\% respectively;  the QCD corrections to the associated
$t\bar{t}H$ production are still not known. The [two--loop] QCD corrections to
the main production mechanism, $gg \ra H$, have been computed
recently and have been found to be rather large since they
increase the cross sections by a factor $\simeq 1.8$ at LHC [there is,
however, an uncertainty of $\sim 20\%$ due to the arbitrariness of the
choice of the renormalization and factorization scales and also of
the parton densities].

The signals which are best suited to identify the produced Higgs particles
at the LHC have been studied in great detail in Ref.~[5,6]. I briefly
summarize here the main conclusions of these studies.

For Higgs bosons in the ``high mass" region, $M_H>140$~GeV, the signal
consists of the so--called ``gold--plated" events $H \ra Z Z^{(*)} \ra 4l^\pm$
with $l=e,\mu$. The backgrounds are relatively small, and one can probe Higgs
masses up to ${\cal O}$(1~TeV) with a luminosity $\int {\cal L}= 100
$~fb$^{-1}$ at LHC. The $H \ra WW^{(*)}$ decay channel is more difficult
to use because of the large background from $t\bar{t}$ pair production.

For the ``low mass" range, the situation is more complicated. The branching
ratio for $H\ra ZZ^*$ becomes too small and due to the huge QCD jet background,
the dominant mode $H\ra b\bar{b}$ is useless; one has then to rely on the rare
$\gamma \gamma$ decay mode with a branching ratio of ${\cal O}(10^{-3})$. At
LHC with a luminosity of $\int {\cal L}= 100$~fb$^{-1}$, the cross section
times the branching ratio leads to ${\cal O}(10^{3})$ events but one has to
fight against formidable backgrounds. Jets faking photons need a rejection
factor larger than $10^{8}$ to be reduced to the level of the physical
background $q\bar{q}, gg \ra \gamma \gamma$ which is still very large. However,
if very good geometric resolution and stringent isolation criteria, combined
with excellent electromagnetic energy resolution to detect the narrow $\gamma
\gamma$ peak of the Higgs boson are available [one also needs a high
luminosity $ {\cal L} \simeq 10^{34}$~cm$^{-2}$s$^{-1}$], this channel,
although very difficult, is feasible. A complementary channel would be the $q
\bar{q} \ra WH, t\bar{t}H \ra \gamma \gamma l \nu$ for which the backgrounds
are much smaller; however the signal cross sections are also very small
making this process also difficult. The process $pp \rightarrow t\bar{t}H$
with the Higgs decaying into $b\bar{b}$ pairs seems also promising provided
that good micro--vertexing to tag the $b$--quarks can be achieved.

In the MSSM, the situation is even more difficult. The production
mechanisms of the SUSY Higgs bosons are practically the same as those of the SM
Higgs; one only has to take the $b$ quark [whose couplings are strongly
enhanced for large $\tb$ values] contributions into account in process $(a)$
[extra contributions from squarks decouple for high masses] and $(d)$ [for
$\tb\gg1$, $q\bar{q}q \ra b\bar{b}H$ becomes the dominant production process].
The various signals for the SUSY Higgs bosons can be summarized as:

i) Since the lightest Higgs boson mass is always smaller than $\sim 140 $
GeV, the $ZZ$ signal cannot be used. Furthermore, the $hWW(h\bar{b}b)$ coupling
is suppressed (enhanced) leading to a smaller $\gamma \gamma$ branching ratio
than in the SM [additional contributions from chargino and sfermion loops can
also alter the decay width] making the search more difficult. If $M_h$ is
close to its maximum value, $h$ has SM like couplings and the situation is
similar to the SM case with $M_H \sim $ 100--140 GeV.

ii) Since $A$ has no tree--level couplings to gauge
bosons and since the couplings of the heavy CP--even $H$ are strongly
suppressed,
the
gold--plated $ZZ$ signal is lost [for $H$ it survives only for small $\tb$ and
$M_H$ values, provided that $M_H<2m_t$]. One has therefore to rely on the $A,H
\ra \tau^+\tau^-$ channels for large $\tb$ values. This mode, which is hopeless
for the SM Higgs, seems to be feasible in this case.

iii) Charged Higgs particles, if lighter than the top quark, can be
accessible in top decays $t \ra H^+b$. This results in a surplus of $\tau$
lepton final states [the main decay mode is $H^-\ra \tau \nu_\tau$] over
$\mu,e$ final states, an apparent breaking of $\tau \ vs. \ e,\mu$
universality.
At LHC, $H^\pm$ masses up to $\sim 100$ GeV can be probed for $m_t \sim 150$
GeV.

Thus, the search for SUSY Higgs bosons is more difficult than the search
for the SM Higgs. Detailed analyses have shown that there is a substantial area
in the SUSY parameter space where no Higgs particle can be found at hadron
colliders. \\

{\bf\noindent 4. Production at e$^+$e$^-$ Colliders}
\vglue 0.2cm
\baselineskip=14pt

At $\ee$ linear colliders operating in the 500 GeV energy range  the
main production mechanisms for Higgs particles are,$^{7,8}$
\begin{eqnarray}
\begin{array}{lccl}
(a)  & \ \ {\rm bremsstrahlung \ process} & \ \ \ee & \ra (Z) \ra Z+H \non \\
(b)  & \ \ WW \ {\rm fusion \ process} & \ \ \ee & \ra \bar{\nu} \ \nu \
(WW) \ra \bar{\nu} \ \nu \ + H \non \\
(c)  & \ \ ZZ \ {\rm fusion \ process} & \ \ \ee & \ra e^+ e^- (ZZ) \ra
e^+ e^- + H \non \\
(d)  & \ \ {\rm radiation~off~tops} & \ \ \ee & \ra (\gamma,Z) \ra
t \bar{t}+H \non
\end{array}
\end{eqnarray}

The bremsstrahlung process is dominant for moderate values of $M_H/\sqrt{s}$,
but falls off like $\sim s^{-1}$ at high energies. The $WW$ fusion
process, on the other hand, is more important for small values of the ratio
$M_H/\sqrt{s}$, i.e. high energies where the cross section grows $\sim
M_W^{-2}$log$s$. For Higgs masses arround 150 GeV, the cross sections for the
bremsstrahlung and $WW$ fusion processes are of comparable size at
$\sqrt{s}=500$ GeV, while the $ZZ$ fusion cross section is smaller by an order
of magnitude. With $\sigma \sim $ 100 fb, a rate of $\sim $ 1000
Higgs particles can be produced at a luminosity $\int {\cal L} = 10 \ {\rm
fb}^{-1}$. For $M_H<120$~GeV, the cross section for $\ee \ra t\bar{t} H$
is of the order of a few fb; this process can be used only  to
measure the $Ht\bar{t}$ Yukawa coupling once the Higgs boson is found.
[Additional production mechanisms are provided by the processes $\gamma \gamma
\ra H $ and $e \gamma \ra \nu WH$, the high energy photons generated by Compton
back--scattering of laser light].

A large variety of channels can be exploited to search for Higgs particles
in the bremsstrah\-lung and fusion processes.$^{7,8}$ In the bremsstrahlung
process, missing--mass techniques can be applied in events with leptonic $Z$
decays and the Higgs may be reconstructed in $H \ra b \bar{b}$ directly.
[Missing--mass techniques of course call for collider designs in which
beamstrahlung is minimal.] These techniques can be applied for Higgs masses $>
100$ GeV; below this value, $b$ tagging through micro--vertex detectors
must be used to separate the $ZH$ signal from the $ZZ$ background. The $WW$
fusion process requires the reconstruction of the Higgs particle, while
missing--mass techniques can also be used in $ZZ$ fusion. Background events
from single $W$ and $Z$ production restrict these experimental search
techniques in the fusion channels to Higgs masses above 100 GeV.

Once the Higgs boson has been found, it will be very important to explore
its properties. This is possible at great detail in the clean environment of
$\ee$ colliders. The zero--spin of the Higgs particle is reflected in the
angular distribution in the bremsstrahl process which asymptotically must
follow the $\sin ^2\theta$ law, corresponding to the predominantly longitudinal
polarization of the accompanying $Z$ boson. Of great importance is the
measurement of the couplings to gauge bosons and matter particles. The strength
of the couplings to $Z$ and $W$ bosons is reflected in the magnitude of the
production cross sections. The relative strength of the couplings to fermions
is accessible through the decay branching ratios. The absolute magnitude is
difficult to measure directly; the measurement is possible only in a small mass
window where the Higgs decays to $b\bar{b}$ and $WW^*$ compete with each other.
Indirect access to the top--Higgs coupling is provided by top--loop mediated
photonic and gluonic Higgs decays [and the reverse processes at $\gamma \gamma$
and pp colliders]. Else, Higgs bremsstrahlung off top quarks or Higgs decays to
$t\bar{t}$ pairs offer opportunities to measure the $ttH$ coupling directly in
limited ranges of the Higgs mass.

An even stronger case for $\ee$ colliders in the 300--500 GeV energy range
is made by the MSSM. In $\ee$ collisons, besides the usual
bremsstrahlung and fusion processes for $h$ and $H$ production, the neutral
Higgs particles can also be produced pairwise: $\ee \ra A + h/H$.
The cross sections for the bremsstrahlung and the pair production as
well as the cross sections for the production of $h$ and $H$ are mutually
complementary, coming either with a coefficient $\sin^2(\beta-
\alpha)$ or $\cos^2(\beta -\alpha)$. The cross section for $hZ$ production is
large for large values of $M_h$, being of ${\cal O}(50$ fb); by contrast, the
cross section for $HZ$ is large for light $h$ [implying small $M_H$].  In major
parts of the parameter space, the signals consist of a $Z$ boson and a
$b\bar{b}$ or a $\tau^+ \tau^-$ pair, which is easy to separate from the main
background, $\ee \ra ZZ$ [for $M_h \simeq M_Z$, efficient $b$ detection is
needed]. For the associated production, the situation is opposite: the cross
section for $Ah$ is large for light $h$ whereas $AH$ production is preferred in
the complementary region.  The signals consists mostly of four $b$ quarks in
the final state, requiring efficient $b\bar{b}$ quark tagging; mass constraints
help to eliminate the QCD jets and $ZZ$ backgrounds. The CP--even Higgs
particles can also be searched for in the $WW$ and $ZZ$ fusion mechanisms.

In $\ee$ collisions, charged Higgs bosons can be produced pairwise, $\ee \ra
H^+H^-$ through $\gamma,Z$ exchange. The cross section depends only on the
charged Higgs mass; it is large up to $M_{H^\pm} \sim 230$~GeV. Charged Higgs
bosons can also be created in laser--photon collisions, $\gamma \gamma \ra
H^+H^-$ but due to the reduced energy, only smaller masses than previously can
be probed. The cross section, however, is enhanced in the low mass range.
Finally, charged Higgs bosons can be produced in top decays as discussed above
in the case of proton colliders. In the range $ 1 < \tb < m_t/m_b$, the $t \ra
H^+b$ branching ratio varies between $ \sim 2 \%$ and $20 \%$ and since the
cross section for $t\bar{t}$ production is ${\cal O}(0.5$~pb) at $\sqrt{s}=500$
GeV, this corresponds to 200 and 2000 charged Higgs bosons at a luminosity
$\int {\cal L} =$ 10 fb$^{-1}$.

The preceding discussion on the MSSM Higgs sector in $\ee$ linear colliders
can be summarized in the following points:
i) The Higgs boson $h$ can be detected in the entire range of the MSSM
parameter space, either through the bremsstrahlung process or through pair
production. In fact, this conclusion holds true even at a c.m.
energy of 300 GeV.
ii) There is a substantial area of	the ($ M_h,\tb$) parameter space where
{\it all} SUSY Higgs bosons can be discovered at a 500 GeV collider.
This is possible if the $H,A$ and $H^{\pm}$ masses are less than $\sim 230$
GeV.
iii) In some parts of the MSSM parameter space, the lightest Higgs $h$ can
be detected, but it cannot be distinguished from the SM Higgs boson. In this
case, Higgs production in $\gamma \gamma$ fusion [which receives extra
contributions from SUSY particles] can be helpful. \\

{\bf\noindent 5. Summary}
\vglue 0.2cm
\baselineskip=14pt

At the hadron collider LHC, the Standard Model Higgs boson can, in
principle, be discovered up to masses of ${\cal O}(1$~TeV). While the region
$M_H>140$
GeV can be easily probed through the $H \ra 4l^\pm$ channel, the $M_H<140$ GeV
region is difficult to explore and a dedicated detector as well as a
high--luminosity is required to isolate the $H\ra \gamma \gamma$ decay.
SUSY Higgs bosons are more difficult to search for and in some areas of the
MSSM parameter space, it is unfortunately possible that no Higgs particle
will be found.

$\ee$ linear colliders with energies in the range of $\sim 500$ GeV are
ideal instruments to search for Higgs particles in the mass range below $\sim
250$ GeV. The search for the Standard Model Higgs particle can be carried out
in several channels and the clean environment of the colliders allows to
investigate thoroughly its properties. In the MSSM, at least the lightest
neutral Higgs particle must be discovered and the heavy neutral and charged
Higgs particles can be observed if their masses are smaller than $\sim$ 230
GeV.
High energy $\ee$ colliders are therefore complementary to hadron colliders.

\vglue 0.6cm
{\bf\noindent References \hfil}
\vglue 0.4cm

\end{document}